# The Evolution of Real-time Remote Intraoperative Neurophysiological Monitoring (IONM)

Jeffrey Balzer[1], Julia Caviness[2], Don Krieger[1]

[1]Department of Neurological Surgery, University of Pittsburgh
[2]Cadwell Industries (Kennewick, WA)

Corresponding Author:
    Don Krieger, PhD
    Department of Neurological Surgery
    Suite 400B Presbyterian University Hospital
    200 Lothrop Street
    Pittsburgh, PA 15232
    kriegerd@upmc.edu
    +1(412)648-9654 office
    +1(412)521-4431 cell

## Abstract

Real-time monitoring of nervous system function with immediate communication of relevant information to the surgeon enables prevention and/or mitigation of iatrogenic injury in many surgical procedures. The hardware and software infrastructure and demonstrated usefulness of telemedicine in support of IONM originated in a busy university health center environment and then spread widely as comparable functional capabilities were added by commercial equipment manufacturers. The earliest implementations included primitive data archival and case documentation capabilities and relied primarily on deidentification for security. They emphasized full-featured control of the real-time data display by remote observers. Today, remote IONM is routinely utilized in more than 200,000 high-risk surgical procedures/year in the United States. For many cases, remote observers rely on screen capture to view the data as it is displayed in the remote operating room while providing sophisticated security capabilities and data archival and standardized metadata and case documentation.

## Introduction

Intraoperative neurophysiological monitoring (IONM) is an ancillary service used in a wide variety of surgical procedures which pose risk to the patient's nervous system. Typically baseline IONM recordings are obtained and recorded from the at-risk neural structure(s) either preoperatively or during surgery before the risk(s) occur. Most commonly these are obtained in the operating room (OR) after induction of anesthesia but before positioning the patient for surgery. Recording continues throughout the procedure and each subsequent recording is compared with both the preceding recordings and with the baseline. When changes are recognized, their likely cause is interpreted, documented, and reported to the surgeon when they represent potential compromise or injury to a neural structure.

The neurophysiological signals are most commonly neuroelectric waveforms. Although machine algorithms may be used to identify time variations in these waveforms [1], the responsibility for those judgements and the assessment of both the cause and the resultant imperative to inform the surgeon is carried by a highly trained and experienced oversight neurophysiologist. Qualified individuals for this integral role in the surgical team are and always have been both essential and rare. The advent and use of telemedicine has been driven by the need to maximize the case coverage provided by each such individual.

**Use of Telemedicine in IONM**

The development of telemedicine support for IONM initially took place at the University of Pittsburgh Medical Center. It was driven by a high and rapidly growing case load. By 1987, the IONM service there was monitoring 1500 cases/year. In addition to two clinical staff neurophysiologists, the service at that time employed several trained and experienced technologists. The technologists set up and broke down equipment in the operating room, attached electrodes to the patients, continuously surveilled the IONM responses, called for the oversight clinical staff as the critical parts of the case approached, communicated with the surgeon when needed, and documented the case. The equipment had been assembled at the university using commercially available components and integrated by an in-house software effort. [2]

At the time, the OR suites were located in four hospitals adjacent to each other with inside corridor access from one to the other. Nevertheless, without remote access to the data, IONM could be compromised when more than two cases required attention simultaneously. In addition, typically 60+ hours/week effort was required from both clinical staff members. To address these problems, ethernet local-area network (LAN) connectivity was installed in the ORs and IONM offices in all four hospitals and a software package was developed and deployed which enabled real-time remote display of up to 16 neurophysiological modalities at a time on any computer on the LAN [2,3].

As an additional burden on the limitations of the technology, many of the surgeons who used IONM services in the university hospitals requested IONM for their procedures in nearby community hospitals. In addition, several surgeons who relied heavily on IONM took jobs elsewhere. In response to these demands, a wide-area-network (WAN) capable data transport layer was added to the remote display package [4,5].

That early work served as proof of principle and as functional specification for both the use of remote capability in IONM and for the underlying technology. Within a decade, commercial IONM devices with remote display capabilities became available and were in wide use.

In 2009, the American Clinical Neurophysiology Society published recommended standards for IONM which included the following statement: *"... The monitoring physician ... is responsible for real-time interpretation of IONM data ... should be present in the operating room or have access to IONM data in real-time from a remote location and be in communication with the staff in the operating room."* [6,7]. Since then, the demand for IONM telemedicine-mediated oversight and interpretation in the United States has grown to 200,000+ cases per year.

## Methods

**Original LAN Remote Display**

As stated above, the initial implementation of remote real-time display in support of IONM extended only over the local area network (LAN) spanning four adjacent university hospitals [3]. The data acquisition computers used in the operating rooms were diskless Apollo workstations (Chelmsford, MA). These machines included a bit-mapped grayscale monitor suitable for waveform display and an ISA bus. An ISA analogue interface board was used to digitize the patient's neurophysiological signals and also provided triggers for physiological stimulators. The processor in the machine was the Motorola 68020 with 68881 floating point unit capable of 70,000 FLOPS. The operating system was Apollo's proprietary Unix-like AEGIS OS which included a fully capable 2-D graphics package [8]. Diskless machines were used to eliminate the risk of disk failure since the machines were routinely moved from room to room.

Network connectivity was provided by a 12 Mbit/sec token ring interface which in practice was more than double the speed of 10 Mbit/sec ethernet. Server machines were positioned strategically in the OR suites and IONM office spaces to support the diskless machines in the ORs and on the oversight neurophysiologists' desks. Each server was configured as a router with both a local token ring interface and an ethernet interface for longer-range communication with the other servers.

The most commonly used neurophysiological modalities were averaged neuroelectric responses to evoked sensory stimuli. Repeated momentary stimuli were presented to the ear (click), eye (flash), or to a peripheral nerve in the arm or leg (electrical stimulus). With each stimulus, a time-locked data segment was digitized and added to the running average of the preceding responses. Ideally, the signal-to-noise ratio progressively improved with $\sqrt{\#stimuli}$ and was consistently interpretable within 30-60 seconds. At the end of an average, the resultant digitized waveform(s) were saved to an archive file along with a time stamp and an annotation when appropriate.

A significant percentage of cases also required neuromuscular monitoring techniques, electromyography (EMG). For these, recording electrodes were placed over muscle groups on the face and/or over the eyes for intracranial tumor resections and pediatric mastoidectomies, and on the arms and legs for spinal fixation case. EMG was monitored and recorded continuously. The signals could be amplified and played in real time over a speaker in the OR and most surgeons preferred having this direct access to the monitored muscle groups in which case they interpreted the signals themselves. In those cases, the signals were still digitized, archived, and interpreted by the staff neurophysiologist as a backup to the surgeon, but interactions mediated by remote display were typically not helpful.

For the Apollo system, each set of waveforms was saved in a 9 x 512-byte record, i.e. 4608 bytes. The first block was used to store stimulus rate, amplifier settings, time stamp, etc. The waveforms were stored in the remaining eight blocks as 4-byte floating point numbers. This compact data structure was a legacy of the original IONM data acquisition system built around Digital Equipment Corporation's PDP-11 computer (Maynard, MA). These machines were limited to a 64 Kbyte program space, were configured with a 20 Mbyte disk, and were rack mounted in closets in the OR suites.

The use of 4608-byte records markedly limited the network demands for remote monitoring. Data access by remote display clients was handled simply by reading the data files for the currently active cases. The routine could also display archived cases. These capabilities were enabled by the LAN-wide file system provided by the AEGIS OS. The programming demands to implement the remote display required only this file access and the 2-D graphics [8] needed to display the waveforms.

The first iteration of the remote display routine required a modest effort to write and deploy. Each modality was displayed in its own window. Each of those windows was initially controlled by a separate instance of the display routine. That awkward and inefficient approach was eliminated in a new software version, rdraw16, which controlled up to 16 windows at a time.

rdraw16 included controls for each window for the range of data epochs to display, the gain to be applied to each channel, and on/off and parameter controls for a variety of signal enhancement algorithms, many of which were developed in-house [5]. In its default mode, rdraw16 continuously polled for new epochs, displayed them when they appeared, scrolled the display to show the most recent 10 epochs in a waterfall display along with two baseline epochs, and showed time stamps and annotations (Figure 1). For each evoked potential modality, a second window could be opened in which the ongoing partially averaged data was displayed every 2-3 seconds. This useful capability was not enabled in the WAN version of the software discussed below.

Insert Figure 1

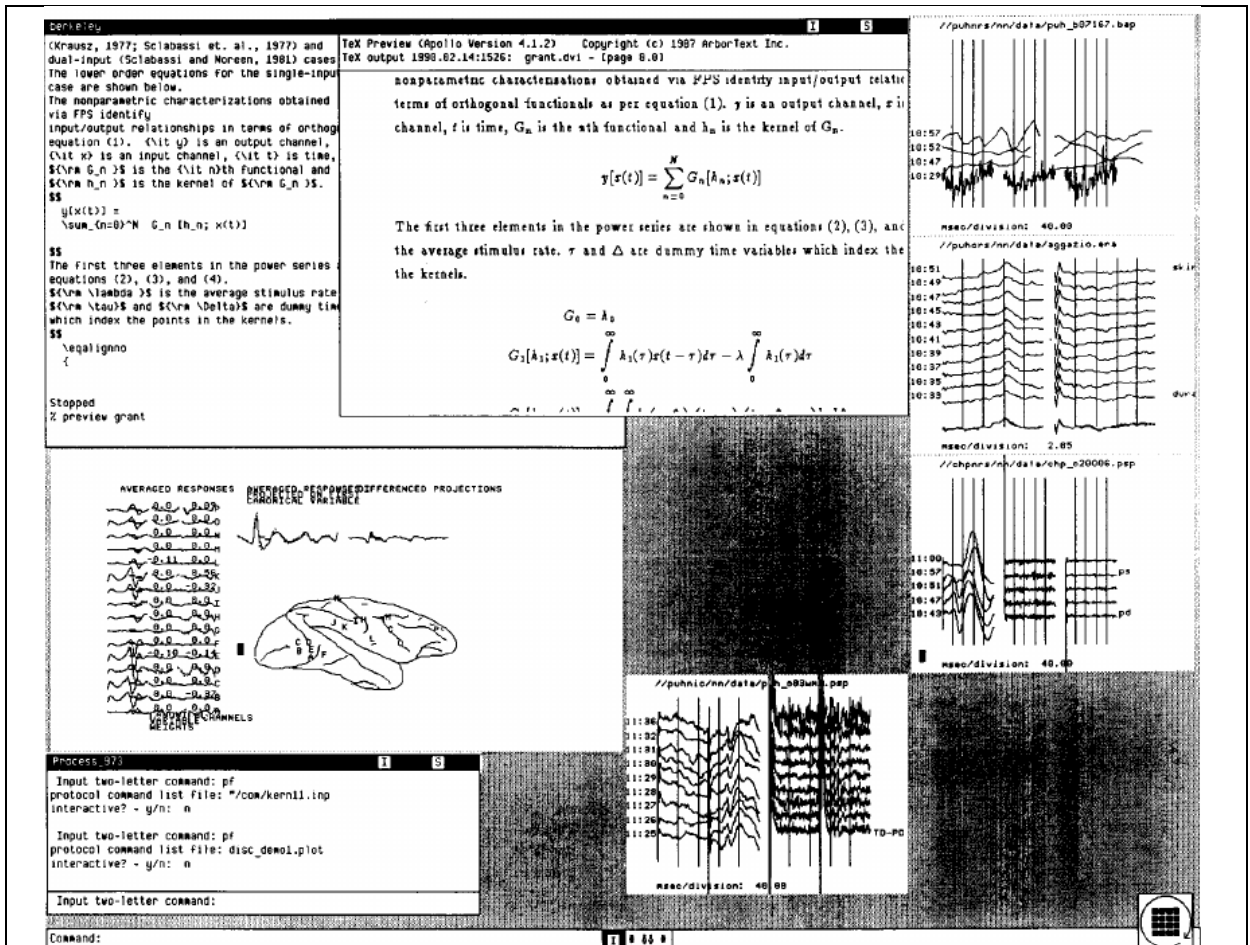

Figure 1. Typical remote monitoring display. This figure shows a display obtainable from anywhere on the LAN. The icon at the lower right corner contains a process that is automatically and continuously searching for active operating room or diagnostic studies. The four small windows containing waterfall displays were spawned by this process and display one ongoing diagnostic study (upper right corner) and three OR cases. The two text windows at the top show a manuscript in preparation. The figure in the center was produced using the process shown in the lower left window. Figure reproduced courtesy of IEEE Computer [3].

A command-line text interface controlled the application and, by default, was minimized to an icon as seen at the lower right corner of the figure. rdraw16 was slaved to a file polling script, rdraw.csh, which intermittently searched for new active files and instructed rdraw16 to spawn a new window with default display when one was found. If no new epoch had been saved to a file for 60 minutes, rdraw.csh instructed rdraw16 to kill that window. Hence this system maintained a real-time display of all active cases automatically and could be left running indefinitely. rdraw.csh controlled rdraw16 through a scratch text file into which it wrote command line instructions which rdraw16 polled, read, and deleted.

Remote display capability was also enabled over standard telephone line 19,200-baud modems. Using a Tektronix-capable graphics monitor [9], an oversight neurophysiologist could dial in to the hospital network, run an rdraw16 instance from the command line and see the display for one modality at a time using the Tektronix 4010 pen plotting graphics layer built into the software. This capability was routinely used from home to provide remote oversight for a case in the hospital and occasionally from the office to follow cases at non-university hospitals which were staffed by university surgeons.

**Original WAN Remote Display**
At the time the original WAN remote display capability (Figure 2) was added [4,5], the underlying equipment had changed and improved again. The machines used in the ORs and offices were Hewlett Packard workstations (Palo Alto, CA) with Motorola 68040 processors (3 MFLOPS), 1-GByte disks, 100 Mbit/sec ethernet interfaces, and a fully capable Unix operating system (HPUX) including X-Windows. The restrictions on data length had been relaxed in the acquisition software package and the case load had grown to 2500+ cases/year. However, the substantive increases in hardware capability and network speed more than made up for the increased demands on the system.

X-windows includes both screen and window capture functionality which were tried for remote display. This approach proved to be unsatisfactory for several reasons. It was awkward to enable and was both slow and negatively impacted the performance of the machine in the OR. It did not enable the neurophysiologist to control his/her waveform display and it necessitated a substantive security risk to the OR machine since it required opening the standard X11 tcp port in the hospital firewall. In recent decades, the performance and security problems with this conceptual approach have been solved and screen capture for remote display is commonly used today in IONM, even though it does not enable control of the displayed waveforms by the oversight neurophysiologist.

For rdraw16, a WAN-capable data transmission layer was added to handle the movement of digitized waveforms across the internet. Parallel Virtual Machine (PVM) [10] and Message Passing Interface (MPI) [11] were considered. Both packages had been developed to support master-slave cluster and supercomputing applications which require more processors than can be placed on a single shared-memory bus. MPI has since become the standard for high performance computing. We selected PVM over MPI because of its open architecture and because Hewlett-Packard and others were providing manufacturer-specific versions of MPI which raised concerns that interoperability might become a problem between machines from different manufacturers.

A separate PVM daemon ran on each machine and mediated message passing and naming services. When a machine was booted in the operating room, its presence on the network was detected by a single polling process running inside the University of Pittsburgh network. Each machine was provided with a fixed IP address and automatically ran an sshd daemon on a nonstandard tcp port. The sshd on that port of each IP address was polled with SYN packets every few minutes using utility nmap. If present, the PVM daemon and an information services daemon (ISD) were spawned on the newly connected machine using ssh. The ISD was developed in-house. It polled the local machine for active data files, provided notifications of their presence across the network, and serviced data requests from rdraw16 instances anywhere on the WAN for data from those files.

Insert Figure 2

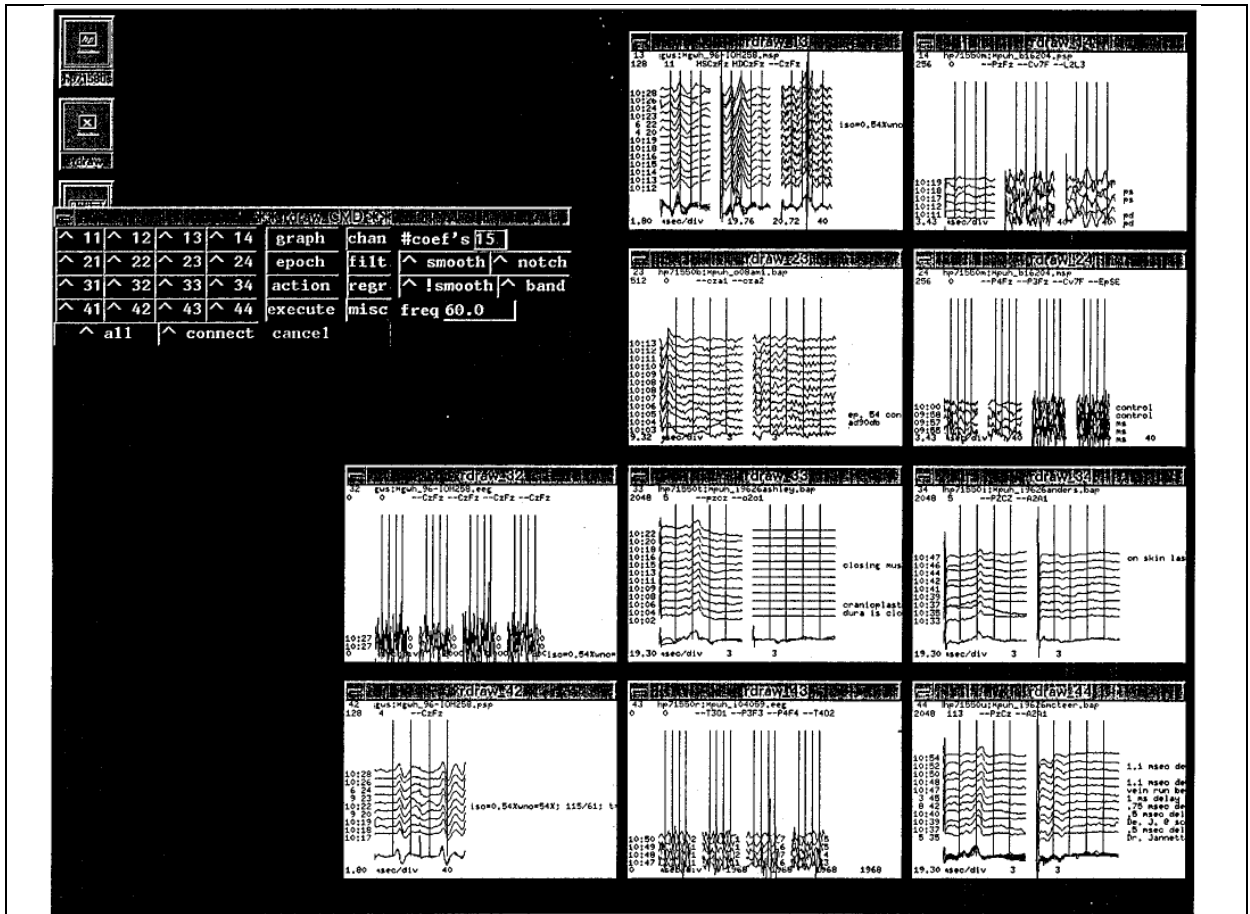

Figure 2. A typical screen image is shown from a workstation with rdraw16 remote monitoring display. Responses are currently shown from seven cases running simultaneously in two hospitals, one of which is in Pittsburgh „puh… and one of which is in Washington, D. C. „gwh…. Each time a recording is completed, the trace appears in the remote display as well as in the operating room. Each waterfall display window includes: „1… a window name, e.g. 24; „2… a name, e.g., hp71550m:*puh_b16204.msp, indicating the name of the computer acquiring the data „hp71550m…, the name of the hospital „puh: Presbyterian University Hospital of Pittsburgh…, the name of the case „b16204…, and the type of responses displayed „bap: brainstem auditory evoked potentials…; „3… the names of the recording channels „International 10/20 System placements P4/Fz, P3/Fz, Cervical C7/Fz, Erb's point…; „4… a series of waveforms with annotations entered by the technician. The window at the upper left labelled **rdraw-CMD** provides control over the waterfall display windows using the mouse and keyboard. This control window was developed using the Tk Toolkit „see Ref. 10…. The iconified window labelled rdraw is a text-only control window. „1… Windows 44, 34, 33, and 23 show brainstem auditory EPs from four different cases in which a craniectomy and decompression of the trigeminal nerve are being performed. „2… Windows 24 and 14 show bilateral median and peroneal nerve SEPs, respectively, from a preoperative diagnostic study in anticipation of spine surgery. „3… Window 43 shows four channel EEG monitoring from a patient with a head injury in an intensive care unit. Windows 13, 42, and 32 show bilateral median nerve, bilateral peroneal nerve, and EEG, respectively, from a craniotomy and resection of a tumor. Figure reproduced courtesy of Annals of Biomedical Engineering [5].

By default, PVM-enabled daemons used tcp sockets to pass messages one to the other and each process-to-process connection required an open file descriptor at each end. This did not scale sufficiently to handle our typical connectivity requirements between rdraw16 instances, ISD's, and PVM daemons. PVM did, however, include udp capability which required only a single file descriptor for each process. We opted to use this and modified our PVM installation to use nonstandard ports for security purposes.

**Commercial Remote Display Dapabilities**

As stated above, the original efforts at the University of Pittsburgh served as a functional specification for the remote display capabilities which followed, most notably for commercially available machines. In 1994, Cadwell Industries (Kennewick, WA) introduced remote display capabilities in its Excel product line. Display capture and transmit technology was used every 15 seconds under Microsoft Windows. Up to nine remote displays could be viewed from monitored cases on the same LAN, four simultaneously (see Figure 3).

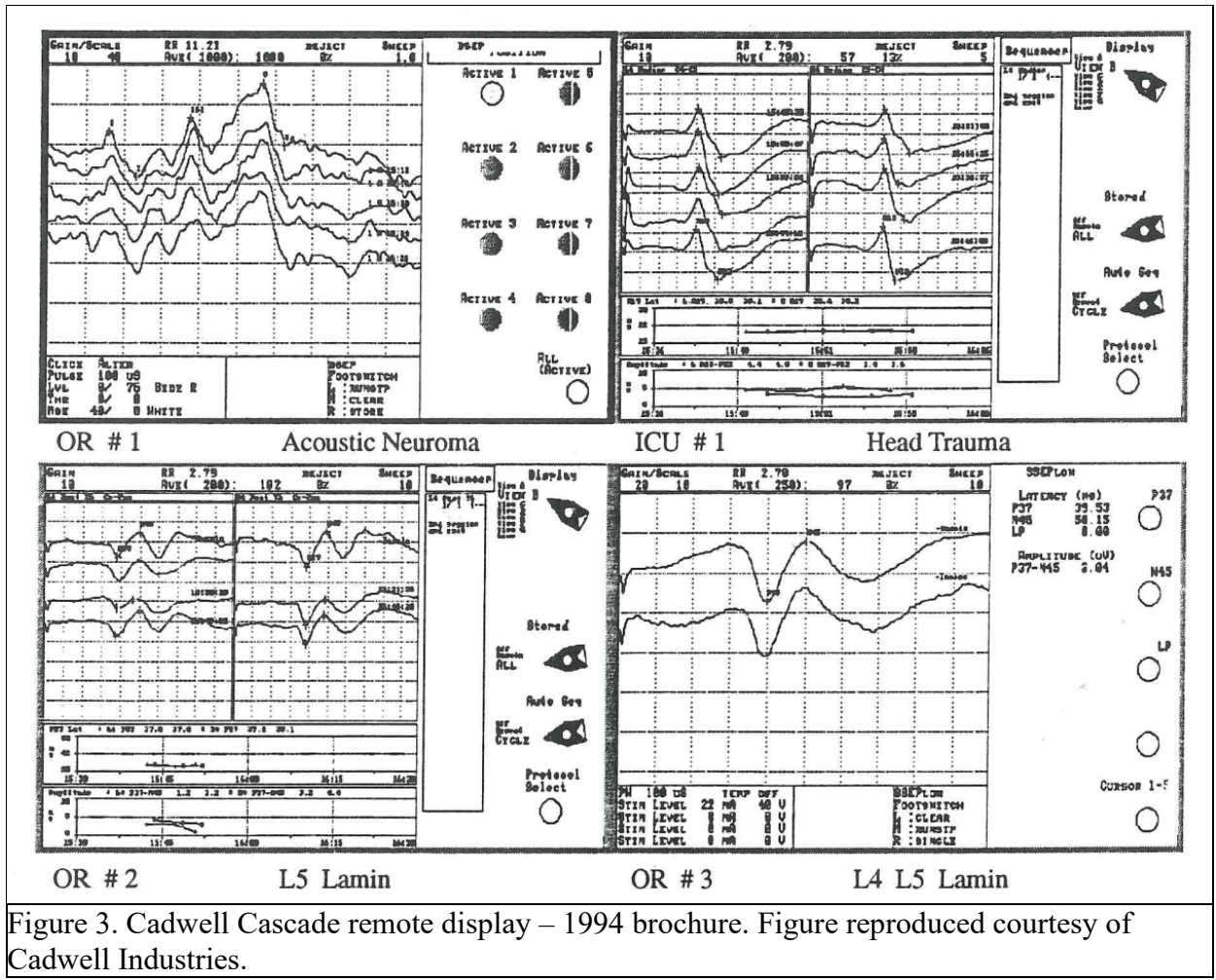

Figure 3. Cadwell Cascade remote display – 1994 brochure. Figure reproduced courtesy of Cadwell Industries.

In 2002, Cadwell introduced "Cascade Classic" software. Their "Remote Reader" enabled simultaneous viewing of up to three waveform sets and included two-way chat, live video, and live EMG. The remote viewer could control her/his data display including channel scaling and which epochs to display, and could view previously recorded cases, all without affecting the data acquisition machine's functions. These full-featured capabilities were usable over the internet. In 2010, Medtronic Xomed's NetOp for Windows (Memphis, TN) provided remote monitoring via LAN or Internet using screen capture. By 2015, they had included remote viewer control of their data display. The Natus Xltek Protektor32 system (Oakville, Ontario, Canada) included remote viewer control of their data display as of 2020 but that is now being phased out. Several manufacturers continue to supply equipment and software capable of multiple-case remote display with full-featured data exploration and display capability by the oversight neurophysiologist.

**Security**

The Healthcare Insurance Portability and Accountability Act (HIPAA) became US federal law in August 1996 [12], but the code sets [13] and privacy rule [14] created by the US Department of Health and Human Services (HHS) under the law did not go into effect until April, 2003. In that interim without benefit of published rules, hospital information technology (IT) managers were understandably over-cautious in handling requests for movement of patient data out of their networks.

With negotiation and with growing experience and understanding of the protections and limitations of newly installed firewalled routers, it became possible and then routine for University of Pittsburgh oversight neurophysiologists to provide IONM services to remote hospitals. Identifying patient health information was excluded from all transmissions across the network. IT managers at remote hospitals placed exceptions in their firewalls for a nonstandard tcp port for ssh to enable spawning pvmd3 on a newly connected OR node and both a primary and two alternate nonstandard udp ports for pvmd3 message transfers.

## Results and Discussion

At every stage of its evolution, IONM has been driven by the demands of surgeons performing high-risk cases and by the responsiveness of hospital administrators. In its infancy, IONM was limited by the financial resources and technical capabilities resident at a few academic medical centers. The primary technical efforts were aimed (a) at enabling efficient data transfer from the computers in the ORs to the computers in the offices of the oversight neurophysiologists and (b) real-time data exploration and display capabilities on those office machines. Secondary emphasis was placed on both security and documentation capabilities and on inter-communications between technologist and staff neurophysiologist. For security, heuristics were applied which fit the demands of each hospital's data security group. Documentation was limited to the notes entered into the data record acquired in the operating room by the technologist in attendance with a hand-written note entered in the patient's chart at the end of each case. Inter-communication was handled using a generic chat program or in some cases, by land-line telephone.

It was the efforts of commercial equipment manufacturers and the development of industry wide medical and legal standards which drove advances in both security and documentation. The Health Level Seven (HL7) standard for transfer of clinical and administrative data by health care providers was initially published in 1989 [15]. As mentioned above, the US Congress' HIPPA regulations came into force in 2003. The Distributed Component Object Model (DCOM) for secure communication between software components on networked computers was publicly launched for Windows 95 in 1996 with full documentation availability beginning in 2006 [16].

Documentation includes contemporaneous annotations which ultimately construct the narrative of the case. This is written and time-stamped electronically by the technologist in the operative room. The time-stamped narrative can then be viewed at any time during the procedure by the oversight neurophysiologist so that consideration of physiologic variables, anesthetic variables and the flow of the surgical procedure can be used as part of the interpretation and differential diagnosis. When the case is complete, all saved data should be moved from the local machine to a central server for storage. This allows for the data to be saved for appropriate periods of time and reviewed in total for quality assurance and control purposes or should a question arise concerning changes which occurred during the case.

More sophisticated inter-communication between technologist and clinical neurophysiologist via chat, voice, and video were added by commercial equipment manufacturers as networking and software infrastructure was added by the computing and networking industries. Universal accessibility of cell phones has provided a backup inter-communication capability which is occasionally useful. For most cases, two-way chat is the fastest and most efficient way for the technologist and oversight neurophysiologist to communicate. Moreover, the technologist acts as the interpretive mouthpiece for the oversight neurophysiologist, thus allowing for communication between the oversight neurophysiologist and surgeon (via the technologist). Two-way audio or multi-way conferencing is typically not be needed as the technologist is able to deliver good or bad news concerning the IONM data. That being said, should specific questions arise that can only be answered by the oversight neurophysiologist, two-way audio or conferences may be a more efficient alternative to just a chat log.

The number of IONM cases in the United States has risen to more than 200,000 per year with the great majority handled with remote neurophysiologist oversight. For many cases, the large commercial service providers rely on screen capture technology for remote display rather than on full featured real-time data exploration and display capabilities. For EMG monitoring of many otolaryngological and spinal instrumentation cases, this limitation is not problematic since the data quality is typically quite good and most surgeons can effectively interpret the signals themselves. For some IONM services, data quality and interpretability of evoked potential monitoring depend more heavily on the remote observer and the observer requires more capable data exploration and display capabilities, e.g. open or endovascular approaches to intracranial aneurysms, open or endoscopic approaches to intracranial mass lesions, fixation of spinal fractures and spinal deformities, traumatic hip replacements, carotid endarterectomies, aortic coarctation repairs, surgeries for spinal cord tumors and spinal dysraphism, open surgery for cranial nerve neuropathy.

The use of screen capture display vs full-featured data access and display control is a choice driven by many factors. There are several advantages to full remote-viewer controllable data display. (a) All data is transmitted to the remote viewer so the raw data set exists on the remote viewer's computer or on a central server accessible to the remote viewer. (b) The remote viewer is able to manipulate the data without effecting the user in the operating room including sizing of individual windows, manipulation of history/waterfalls, individual manipulation of data amplitude and time bases as well as minimizing data window(s). (c) In screen capture cases, the technologist is responsible for making sure that data can be viewed remotely by having the data and event logs visible on their desktop as well as being solely responsible for changing gains, time-bases etc should the oversight neurophysiologist request them. This can present a screen "real-estate" issue for both the technologist in the OR and for the oversight neurophysiologist. (d) If personal health information is displayed on the IONM machine in the OR, screen capture and transmission of the window containing it is a HIPPA violation. (e) Data review is often required during or after the procedure to facilitate quality assurance or to resolve questions concerning sensitivity of the IONM. The display of current data along with baseline waveforms from hours or days before is most often viewed in a "stacked" or "waterfall" format which typically requires full featured control to optimize the interpretability of the display. The choice to deploy an IONM system without these capabilities is effectively made by the oversight neurophysiologist in consideration of (a) the case mix for which the system will be used, (b) the work style she/he uses for oversight, and (c) the training and experience of the technologist(s) who will be in the OR.

# References

1. Krieger D, Sclabassi RJS. "Time-varying evoked potentials." J Med Engg Tech 18(3), 1994: 96-100.
2. Krieger DN, Lofink RM, Doyle EL, Burk G, Sclabassi RJ. "Neuronet: Implementation of an Integrated Clinical Neurophysiology System." Medical Instrumentation 21(6), 1987: 296-303.
3. Krieger D, G Burk, RJ Sclabassi. "Neuronet: A distributed real-time system for monitoring neurophysiologic function in the medical environment." IEEE Computer 24(3): 45-55, 3/91.
4. Krieger D, Simon B, Chay T, Sclabassi R. "A WAN Surgical Monitoring Application Built on PVM," Proceedings of the 1995 PVM Users' Group Meeting, Pittsburgh, PA, May 1995. A WAN Surgical Monitoring Application Built on PVM (cmu.edu)
5. Krieger D, Onodipe S, Charles PJ, Sclabassi RJ. "Real Time Signal Processing in the Clinical Setting." Annals of Biomedical Engg (26):462-472, 1998.
6. Nuwer, M.R. Overview and history. Intraoperative monitoring of neural function. M.R. Nuwer. Amsterdam, Elsevier; 2008: 2-6.
7. American Clinical Neurophysiology Society. "Guideline 11A: Recommended standards for neurophysiologic intraoperative monitoring – principles. 2009. Guideline 11A: RECOMMENDED STANDARDS FOR (acns.org)
8. Apollo Computer Inc. "Domain Graphics Primitive Resource Call Reference." Order No. 007194, Rev 02, 1987. 007194-02_Domain_Graphics_Primitive_Resource_Call_Reference_Jun87.pdf (bitsavers.org)
9. Digital Equipment Corporation. "VT240 Series Owner's Manual." 1984:86-88. VT240 Series Owner's Manual (vt100.net)
10. Geist A, Beguelin A, Dongarra J, Jiang Weicheng, Manchek R, Sunderam VS. PVM: A Users' Guide and Tutorial for Network Parallel Computing. The MIT Press, 1994. https://doi.org/10.7551/mitpress/5712.001.0001
11. Clarke L, Gendinning I, Hempel R. The MPI Message Passing Interface Standard. Programming Environments for Massively Parallel Distributed Systems, Monte Verita, Switzerland, 1994, Springer The MPI Message Passing Interface Standard | SpringerLink
12. H.R.3103 - Health Insurance Portability and Accountability Act of 1996. United State Congress. Text of the bill online
13. HIPPA Code Sets Overview. Centers for Medicare and Medicaid Services. Online.
14. The HIPPA Privacy Rule. Health and Human Services. Online.
15. HL7 International Version 2 Product Suite. HL7 documentation online
16. Microsoft [MS-COM]: Distributed Component Object Model (DCOM) Remote Protocol. online